\documentclass[prb,twocolumn,showpacs]{revtex4}
\input epsf
\epsfverbosetrue

\usepackage{graphics}
\usepackage{amsmath}
\usepackage{bm}
\usepackage{epsfig}
\newcommand{\Journal}[4]{#1 \textbf{#2}, #3 (#4)}
\newcommand{\PRev}{Phys. Rev.}

\newcommand{\be}{\begin{eqnarray}}
\newcommand{\ee}{\end{eqnarray}}
\newcommand{\nn}{\nonumber}

\newcommand{\pr}{PrOs$_4$Sb$_{12}$}

\begin{document}
\title{Field angle-dependent thermal conductivity in nodal superconductors}
\author{T. R. Abu Alrub and S. H. Curnoe}
\affiliation{Department of Physics and Physical Oceanography,
Memorial University of Newfoundland, St. John's, NL, A1B 3X7, Canada}

\begin{abstract}
We apply a semi-classical method to the problem of field angle-dependent 
oscillations of the
density of states and thermal conductivity 
for nodal superconductors and
apply our results to the superconductor \pr.   
The oscillatory contributions to the thermal conductivity for all
possible point node configurations for a superconductor with 
$T_h$ symmetry are calculated.   It is found that experimental
results are best accounted for by nodes in the off-axis
directions $[\pm \sin\phi_0, 0, \pm\cos\phi_0]$, which are associated
with the time-reversal breaking, triplet paired phase with symmetry $D_2(E)$. 
\end{abstract}

\pacs{74.20.-z, 71.27.+a, 71.10.-w}
\maketitle
\section{Introduction}
The low temperature thermodynamic properties of unconventional superconductors 
are governed by nodal quasiparticles. 
In the presence of a  small magnetic field 
$H_{c1}\leq H\ll H_{c2}$, it was shown by Volovik \cite{volovik1993} that 
the dominant contribution to 
the density of states (DOS) for superconductors with line nodes
comes from delocalized quasiparticles, 
in contrast to 
$s$-wave superconductors in which the DOS
is dominated by quasiparticles localised inside vortex cores.\cite{Caroli1964,volovik1993}
Volovik argued that the delocalised states 
experience a semi-classical  adjustment to their energy due to the
magnetic field which is expressed as a 
Doppler shift
$\omega\rightarrow\omega-
{\bm v}_s(\bm r)\cdot{\bm k}$, 
where ${\bm v}_s(\bm r)=\frac{1}{2mr}\hat{\beta}$ is the superfluid velocity 
and $\beta$ is the winding angle
around a single vortex. 
As a result, contributions proportional to 
the magnetic field have been predicted to appear in 
thermodynamic and transport properties of line node 
superconductors\cite{volovik1993,Barash1996, Barash1997,Hirschfeld1998,Kubert1998/1,Kubert1998/2,Vekhter1999/1, Franz1999, Vekhter1999/2,Vekhter2000}.  

Magnetic contributions that appear because of the Doppler shift will 
strongly depend on the position of the  
magnetic field with respect to the nodes. 
Consequently, oscillations in the DOS and related quantities
have been predicted for superconductors with line nodes in a 
rotating magnetic field,\cite{volovik,Moler1997,Hirschfeld1998,Vekhter1999/3,Vekhter2001,Won2001,Miranovic2003,Udagawa2004,
Thalmeier2005,Tewordt2005,Vorontsov2006,Vorontsov2007/1,
Vorontsov2007/2} 
and have been observed experimentally in in-plane thermal conductivity 
in YBa$_2$Cu$_3$O$_7$.\cite{Aubin1997,Yu1995} 
Similar results have been found for other
unconventional superconductors,\cite{Izawa2003,Izawa2001/1,Izawa2001/2,Izawa2001/3, Izawa2002, Watanabe2004,Matsuda2006}
including YNi$_2$B$_2$C\cite{Izawa2002} and PrOs$_4$Sb$_{12}$,\cite{Izawa2003}
which are reported to have point nodes instead of line nodes.
Oscillations in the angular field dependence of the specific heat 
have also been observed in  
several unconventional superconductors.\cite{Park2003,Park2004,Deguchi2004/1,Deguchi2004/2,Aoki2004,Custers2006,Custers2007,Sakakibara2007}

Volovik's proof that, in superconductors with line nodes,
delocalised quasi-particles have a greater
contribution to the low-energy DOS than vortex localised
quasi-particles does not extend to superconductors with point nodes.
However, we are interested in finding the oscillatory component of the
DOS in a rotating magnetic field, for which the semi-classical method,
applied to delocalised quasi-particles, may be valid.
For delocalised states, oscillations are obtained just by Doppler-shifting
the quasiparticle energies.  For localised states, the amplitude of
oscillations may be found by calculating separately the DOS for the case 
when the field
points in the direction of the nodes and the case when the field is
perpendicular to the nodes and subtracting the results for the two
cases.  The former case corresponds to the $s$-wave result, which gives
the contribution from localised 
quasi-particles as $N_{s-{\rm loc.}} \sim N_F \xi^2/R^2 \sim N_F H/H_{c2}$,
where $\xi$ is the coherence length and $R$ is the inter-vortex spacing.  This
sets a lower bound on the DOS contributions from localised 
quasi-particles in superconductors with line nodes or point nodes.  
The upper bound of the
DOS contribution from localised quasiparticles in superconductors
with nodes 
is found when the field is parallel to the
nodes.
Volovik found that for superconductors with line nodes, the
localised quasiparticles contribute $N_{\rm line-loc.} \sim N_F \sqrt{H/H_{c2}}/\log\sqrt{H_{c2}/H}$ 
which he found to be less  than  the delocalised contribution
$N_{\rm line-deloc.} \sim N_F \sqrt{H/H_{c2}}$.
The oscillatory contribution for the delocalised states is contained within
$N_{\rm line-deloc.}$, while the oscillation amplitude for the 
localised states is $\sim N_{\rm line-loc.} - N_{s-{\rm loc.}} 
\approx N_{\rm line-loc.}$.
Thus the DOS oscillations in a 
line node superconductor are dominated by the delocalised contribution,
and the semi-classical treatment is valid.  
For a point node superconductor, one finds that the delocalised
contribution is $N_{\rm point-deloc.} \sim N_{\rm point-loc.} \sim 
N_{s-{\rm loc.}}$.
Again, the oscillatory contribution for delocalised states
is contained within  $N_{\rm point-deloc.}$ while the
oscillation amplitude for the localised states is found by comparing
$N_{\rm point-loc.}$ to $N_{s-{\rm loc.}}$.
This suggests that while both localised and delocalised states contribute
to the DOS in point node superconductors,
the oscillatory component is
dominated by the delocalised states.
We will assume that this is the case, but a more thorough
investigation of the role of vortex localised quasiparticles in point node 
superconductors is 
warranted.

Field dependent thermal conductivity measurements are usually performed in 
one of two experimental configurations.  
In  layered compounds in which the c-axis
conductivity is low, such as 
in YBa$_2$Cu$_3$O$_7$, 
the in-plane conductivity is usually measured with the B-field rotating in the
same plane.  
Then one in-plane component of the current will be parallel to
vortices produced by the B-field while the other in-plane component of 
the current is perpendicular to the vortices.
This introduces complications when calculating the different
components of the current averaged over the vortex lattice,
since a different kind of averaging procedure should be used 
depending on whether the heat current is parallel or
perpendicular to the vortices.\cite{Kubert1998/2}
As a consequence, for a field rotating in the $xy$ plane, 
in-plane components of the conductivity will oscillate
with a period of twice the field angle even when there are no
nodes at all, and this oscillation will dominate any 
nodal contribution.\cite{Matsuda2006,Maki1967}
Thus, whenever possible, the preferred set-up is to measure the
heat currents perpendicular to the rotating B-field.\cite{Matsuda2006}

In the superconductor \pr, the pairing symmetry is widely thought to be 
unconventional,\cite{Izawa2003,Maple2001,Bauer2002,Aoki2003,Chia2003,Huxley2004,Frederick2005,Nishiyama2005,Higemoto2007,Katayama2007} 
with spin triplet pairing\cite{Higemoto2007}
and broken time reversal symmetry.\cite{Aoki2003}  Power law behavior 
has been  observed in many thermodynamic and transport measurements at low 
temperature,\cite{Izawa2003,Bauer2002,Frederick2005,Chia2003,Katayama2007} which suggests the 
existence of nodes in the gap function; however  
a nodeless gap function has been observed in some experiments.
\cite{MacLaughlin2002,Kotegawa2003,Suderow2004,Seyfarth2006}
Oscillations of the thermal conductivity in a rotating magnetic field
are another indication that there are nodes in the gap function.\cite{Izawa2003}
In previous works,\cite{Sergienko2004,Tayseer2007/1} we 
have attempted to determine the  symmetry of the 
superconducting state in \pr\ using available experimental results. 
Among the various possible choices, 
we selected the spin triplet paired
states belonging to the three dimensional irreducible representation 
$T_u$ of the point group $T_h$ with symmetry 
$D_2(C_2)\times {\cal K}$ and order parameter components
$(0,0,1)$ and 
$D_2(E)$ with  components $(i|\eta_2|,0, |\eta_1|)$.
We label these phases `A' and `B' respectively. 
The A phase is unitary, and has two cusp point nodes in the directions $\pm[0,0,1]$, while 
the B phase is nonunitary and 
has four cusp point nodes in the directions [$\pm\sin\phi_0$,0,$\pm\cos\phi_0$],
where $\phi_0$ is an angle determined from phenomenological parameters.\cite{Tayseer2007/1} 
We will consider these and all other symmetry-allowed phases with point nodes.

In this article we calculate the DOS and residual transport under
an applied magnetic field.
We consider both the clean and dirty limits, in which the impurity scattering rate is much smaller or greater than the Doppler shift, 
respectively. 
For the purpose of comparison, we
begin by stating in Section II 
results for the residual DOS and thermal conductivity 
for the $d$-wave (line node) superconductors.  
Section III is devoted to point nodes applied to PrOs$_4$Sb$_{12}$.
In Section IV we compare our results to experiment.
Concluding remarks are made in Section V.
 
\section{Density of States and thermal conductivity for 
superconductors with line nodes}
In this section we 
consider a $d$-wave superconductor (such as YBa$_2$Cu$_3$O$_7$)
with line nodes along  the directions $k_x = \pm k_y$ and a magnetic
field applied in the $k_xk_y$-plane at an angle $\epsilon$ with respect
to the $x$ axis.
In the vicinity of a node, the gap function takes the form
$\Delta({\bm k}) \approx v_g k_2$, where $k_2$ points perpendicular to the 
node in the $xy$-plane and ${\bm v}_g = \left.\frac{\partial \Delta({\bm k})}{\partial{\bm k}}
\right|_{\rm node}$ is the gap velocity.  
The quasiparticle energy is 
$E({\bm k}) = \sqrt{\varepsilon^2({\bm k})+\Delta^2({\bm k})} \approx
\sqrt{v_F^2 k_1^2 +v_g^2k_2^2}$, where 
$k_1$ points in the direction of the node.
Thus, in the vicinity of a node, the Green's function takes the form
\be
G(\bm k, i{\tilde\omega}_n,{\bm r})=\frac{i{\tilde\omega}_n+\alpha_j(\bm r)+v_F k_1}
{(i{\tilde\omega}_n+\alpha_j(\bm r))^2+v_F^2k_1^2+v_g^2k_2^2}
\ee
where
$i{\tilde\omega}_n=i\omega_n+i\Gamma_0$,
$\alpha_{j}(\bm r)={\bm v}_s(\bm r)\cdot{\bm k}_{Fj}$ is the Doppler shift at the $j$th node
and $\Gamma_0 = -\Im\Sigma_{ret}(\omega=0)$ is the scattering rate at zero energy. 
The self-energy $\Sigma$ is derived from the T-matrix formalism
for impurity scattering
and is the solution to the self-consistent equation\cite{Pethick1986,Hirschfeld1998}
\be
\label{4.2}
\Sigma(i\omega_n)=\frac{\Gamma G_0(i\tilde{\omega}_n,{\bm r})}{c^2-G_0^2(i\tilde{\omega}_n,{\bm r})}
\ee
where $\Gamma$ is proportional to the impurity concentration,
$c$ is related to the
the phase shift $\delta_0$, $c=\cot\delta_0$
and
\be
G_0(i{\tilde\omega}_n,{\bm r})=\frac{1}{\pi N_F}\sum_{\bm k}G(\bm k, i{\tilde\omega}_n,{\bm r}).
\label{4.1.1}
\ee
$N_F$ is the density of states at the Fermi surface.
In the unitary limit ($c=0$) this yields a self-consistent equation for the
scattering rate\cite{note1,Kubert1998/2}
\be
\Gamma_0^2=  \pi^2 N_F v_F v_g\Gamma\bigg[\ln\left(\frac{p_0^2}{\sqrt{(\alpha_1^2(\bm r)+\Gamma_0^2)(\alpha_2^2(\bm r)+\Gamma_0^2)}}\right)
\nn\\
+\frac{\alpha_1(\bm r)}{\Gamma_0}\tan^{-1}\left(\frac{\alpha_1(\bm r)}{\Gamma_0}\right)+\frac{\alpha_2(\bm r)}{\Gamma_0}\tan^{-1}\left(\frac{\alpha_2(\bm r)}{\Gamma_0}\right)\bigg]^{-1}.
\label{gamma}
\ee
where $p_0$ is a cutoff and $\alpha_{1,2}(\bm r)$ are the Doppler shifts at two opposite nodes and can be written as
\begin{eqnarray}
\alpha_1(\bm r)&=&\frac{k_F}{2m r}\sin{\beta}\sin{(\pi/4-\epsilon)}
\\
\alpha_2(\bm r)&=&\frac{k_F}{2m r}\sin{\beta}\cos{(\pi/4-\epsilon)}
\end{eqnarray}
where $r$ is the distance from the centre of the vortex core, $\beta$ is the vortex winding angle and 
$\epsilon$ is the angle of the magnetic field relative to the $x$-axis.

\subsection{Density of States}
The DOS is given by
\be
-N(\omega,{\bm r})
= \frac{1}{\pi}\int \frac{d^3 k}{(2\pi)^3} 
\Im{G_{ret}({\bm k},\omega,{\bm r})}
\label{dos}
\ee
where the integral over $\bm k$ is evaluated as the sum 
of four separate volume integrations centred about each node.\cite{Durst2000}
Then the DOS at the Fermi energy is
\begin{eqnarray}
N(0,\bm r)
& =& \frac{\Gamma_0}{\pi^2 v_F v_g}\bigg[\ln\left(\frac{p_0^2}{\sqrt{(\alpha_1^2(\bm r)+\Gamma_0^2)(\alpha_2^2(\bm r)+\Gamma_0^2)}}\right)
\nn\\
&&+\frac{\alpha_1(\bm r)}{\Gamma_0}\tan^{-1}\left(\frac{\alpha_1(\bm r)}{\Gamma_0}\right)+\frac{\alpha_2(\bm r)}{\Gamma_0}
\nn\\
&&\times\tan^{-1}\left(\frac{\alpha_2(\bm r)}{\Gamma_0}\right)\bigg]
\end{eqnarray}
In the clean limit $(\Gamma_0/|\alpha(\bm r)|)\rightarrow 0$ and
\begin{eqnarray}
N(0,{\bm r})&\approx&\frac{|\alpha_1(\bm r)|+|\alpha_2(\bm r)|}{2\pi v_F v_g}\\
&=&\frac{k_F}{2m r} |\sin{\beta}|
\frac{\max{[|\sin\epsilon|,|\cos\epsilon|]}}{2\sqrt{2}\pi v_F v_g}.
\end{eqnarray}
This result necessarily has the same form as the finite
frequency DOS $N(\omega) \sim |\omega|$ of 
superconductor with line nodes.
Averaging over the vortex unit cell we obtain the result
as in Ref.\ \onlinecite{volovik},
\be
\left<N(0,{\bm r})\right>_H &=&\frac{1}{\pi R^2}\int_{\xi_0}^Rdr\, r\int_0^{2\pi}d\beta\, {N(0,{\bm r})}
\label{average}
\\
&\sim& N_F\frac{\xi_0}{R}
\max{[|\sin\epsilon|,|\cos\epsilon|]}
\ee
where $\frac{\xi_0}{R} \sim \sqrt{\frac{H}{H_{c2}}}$. 
Evidently, there are four-fold oscillations in the DOS
as a function of the field angle $\epsilon$.

In the dirty limit $|\alpha(\bm r)|/\Gamma_0\ll 1$  we find
\be
N(0,{\bm r})=\frac{\Gamma_0}{\pi^2 v_F v_g}\left[2\ln\left(\frac{p_0}{\Gamma_0}\right)+\frac{\alpha_1^2(\bm r)+
\alpha_2^2(\bm r)}{\Gamma_0^2}\right].
\ee
The first term is just the impurity induced DOS $N(0)$,\cite{Durst2000} so
\be
\delta N(0,{\bm r}) = 
N(0,{\bm r})-N(0)
=\frac{1}{4\pi^2r^2}
\frac{v_F}{v_g \Gamma_0}\sin^2\beta
\ee
and the average DOS is\cite{Kubert1998/1}
\be
\left<\delta N(0,{\bm r})\right>_H\sim
N_F \frac{\Delta_0}{\Gamma_0}\frac{H}{H_{c2}}
\ln\left(\frac{H_{c2}}{H}\right).
\ee
where $\Delta_0=\frac{v_F}{\xi_0}\sim N_F v_F v_g$.
In this case, impurities remove the field directional
dependence of the DOS.

\subsection{Thermal conductivity}
The thermal conductivity tensor is defined by the Kubo formula.\cite{Mahan2000}
In the limit $T \rightarrow 0$ it is expressed in terms of the 
imaginary part of the Green's function as\cite{Durst2000,Tayseer2007/2} 
\begin{equation}
\frac{\widetilde{\kappa}(0,{\bm r})}{T}
=\frac{k_B^2}{3}\sum_{\bm k}{\bm v}_F{\bm v}_F\ {\rm Tr}[\Im\widetilde{G}_{ret}(0,{\bm r})\Im\widetilde{G}_{ret}(0,{\bm r})],
\label{kubo}
\end{equation}
where $k_B$ is the Boltzmann constant and 
${\bm v_F}$ is the Fermi velocity in the direction of ${\bm k}$.
By again dividing the integration over ${\bm k}$ into regions
centred over each node,
this eventually leads to
\be
\frac{\widetilde{\kappa}(0,{\bm r})}{T}=&&
\frac{k_B^2}{3}\frac{\sum_{j=1}^4{\bm v}_F{\bm v}_F}{(2\pi)^2 v_F v_g}\int_0^{2\pi}d\theta \int_0^{p_0}dp\ p
\nn\\
&&\times\frac{2\Gamma_0^2}
{\left[(\alpha_j(\bm r)+p)^2+\Gamma_0^2\right]^2}
\ee
where the integration variable is $p = \sqrt{v_F^2 k_1^2 + v_g^2 k_2^2}$.
Performing the integration yields
\begin{equation}
\frac{\widetilde{\kappa}(0,{\bm r})}{T}=\frac{k_B^2}{3}\frac{\sum_{j=1}^4{\bm v}_F{\bm v}_F}{\pi v_F v_g} 
\bigg(1 + \frac{\alpha_j(\bm r)}{\Gamma_0}
\left[\tan^{-1}\left(\frac{\alpha_j(\bm r)}{\Gamma_0}\right) - \frac{\pi}{2}\right]
\bigg)
\end{equation}
where now ${\bm v}_F$ is evaluated at each node.
Summing over nodes  yields
\begin{widetext}
\be
\frac{\delta\widetilde{\kappa}(0,{\bm r})}{T}&=&\frac{\widetilde{\kappa}(0,{\bm r})-\widetilde{\kappa}(0,0)}{T} =
\frac{k_B^2}{6\pi}\frac{v_F}{v_g}
\nn\\
&&\times\left( \begin{array}{cc}
\frac{\alpha_1(\bm r)}{\Gamma_0}\tan^{-1}\left(\frac{\alpha_1(\bm r)}{\Gamma_0}\right)+\frac{\alpha_2(\bm r)}{\Gamma_0}\tan^{-1}\left(\frac{\alpha_2(\bm r)}{\Gamma_0}\right)& 
\frac{\alpha_1(\bm r)}{\Gamma_0}\tan^{-1}\left(\frac{\alpha_1(\bm r)}{\Gamma_0}\right)-\frac{\alpha_2(\bm r)}{\Gamma_0}\tan^{-1}\left(\frac{\alpha_2(\bm r)}{\Gamma_0}\right)\\
\frac{\alpha_1(\bm r)}{\Gamma_0}\tan^{-1}\left(\frac{\alpha_1(\bm r)}{\Gamma_0}\right)-\frac{\alpha_2(\bm r)}{\Gamma_0}\tan^{-1}\left(\frac{\alpha_2(\bm r)}{\Gamma_0}\right)& 
 \frac{\alpha_1(\bm r)}{\Gamma_0}\tan^{-1}\left(\frac{\alpha_1(\bm r)}{\Gamma_0}\right)+\frac{\alpha_2(\bm r)}{\Gamma_0}\tan^{-1}\left(\frac{\alpha_2(\bm r)}{\Gamma_0}\right)
\end{array} \right)
\ee

In the clean limit $\alpha(\bm r)\gg\Gamma_0$, the thermal conductivity is
\be
\frac{\delta\widetilde{\kappa}(0,{\bm r})}{T}=\frac{k_B^2}{12}\frac{v_F}{v_g}
\left( \begin{array}{cc}
\frac{|\alpha_1(\bm r)|+|\alpha_2(\bm r)|}{\Gamma_0}& 
\frac{|\alpha_1(\bm r)|-|\alpha_2(\bm r)|}{\Gamma_0}\\
\frac{|\alpha_1(\bm r)|-|\alpha_2(\bm r)|}{\Gamma_0}& 
 \frac{|\alpha_1(\bm r)|+|\alpha_2(\bm r)|}{\Gamma_0}
\end{array} \right),
\ee
and the average over the vortex unit cell  is

\be
\left<\frac{\delta\widetilde{\kappa}(0,{\bm r})}{T}\right>_H \sim \ k_B^2\frac{v_F}{v_g}\frac{\Delta_0}{\Gamma_0}\sqrt{\frac{H}{H_{c2}}}
\left( \begin{array}{cc}
\frac{1}{\sqrt{2}}\max[|\sin\epsilon|,|\cos\epsilon|]& 
|\sin{(\pi/4-\epsilon)}|-|\cos{(\pi/4-\epsilon)}|\\
|\sin{(\pi/4-\epsilon)}|-|\cos{(\pi/4-\epsilon)}|& 
 \frac{1}{\sqrt{2}}\max[|\sin\epsilon|,|\cos\epsilon|]
\end{array} \right).
\ee
\end{widetext}

In the dirty limit $\alpha(\bm r)\ll\Gamma_0$ we find
\be
\frac{\delta\widetilde{\kappa}(0,{\bm r})}{T}=\frac{k_B^2}{6\pi}\frac{v_F}{v_g}
\left( \begin{array}{cc}
\frac{\alpha_1^2(\bm r)+\alpha_2^2(\bm r)}{\Gamma_0^2}& 
\frac{\alpha_1^2(\bm r)-\alpha_2^2(\bm r)}{\Gamma_0^2}\\
\frac{\alpha_1^2(\bm r)-\alpha_2^2(\bm r)}{\Gamma_0^2}& 
 \frac{\alpha_1^2(\bm r)+\alpha_2^2(\bm r)}{\Gamma_0^2}
\end{array} \right)
\ee
The average over the vortex unit cell is
\begin{eqnarray}
& & \left<\frac{\delta\widetilde{\kappa}(0,{\bm r})}{T}\right>_H \nonumber \\
& & \sim
k_B^2\frac{v_F}{v_g}\frac{\Delta_0^2}{\Gamma_0^2}\frac{H}{H_{c2}}
\ln\left(\frac{H_{c2}}{H}\right)
\left( \begin{array}{cc}
1&-\sin2\epsilon \\
 -\sin2\epsilon& 1
\end{array} \right).
\end{eqnarray}
Thus, as in the DOS, impurities remove oscillations due to 
nodes in the diagonal components
of the thermal conductivity.

\section{Density of states and thermal conductivity for a point node
 superconductor} 
We will first assume that there are an arbitrary number of
{\em linear} ({\em i.e.} vanishing linearly with momentum) point
nodes in the gap function, and that the gap velocities $v_g$ are equal and
isotropic around each node. 
We begin by finding a self-consistent equation for the
scattering rate $\Gamma_0$ analogous to  Eq.\ \ref{gamma}.
For point nodes, 
Eq.\ \ref{4.1.1} is 
\be 
G_0(0,{\bm r})& =&
\frac{1}{\pi N_F}\frac{\sum_{\rm nodes}}{(2\pi)^3v_F v_g^2}\int_0^{2\pi}d\phi\int_0^{\pi}d\theta \sin{\theta}\int_0^{p_0}dp\ p^2
\nn\\
&&\times\frac{-\alpha_j(\bm r)+i\Gamma_0+p\cos\theta}{(-\alpha_j(\bm r)+i\Gamma_0)^2-p^2}
\label{25}
\ee
where the integration variable is $p  = \sqrt{v_F^2 k_1^2 + v_g^2( k_2^2 + k_3^2)}$
and $k_1$ is parallel to the node while $k_{2,3}$ are perpendicular to the
node.  In  Eq. \ref{25}, we have again divided the volume of
integration into parts each centred around a node.
The integrations yield
\be
G_0(0,{\bm r})&=&\frac{-i}{N_F 2\pi^3v_F v_g^2}\sum_{\rm nodes}
\bigg[(\Gamma_0+i\alpha_j(\bm r))p_0
\nn\\
&&-\frac{\pi}{2}(\Gamma_0+i\alpha_j(\bm r))^2\bigg].
\label{24}
\ee
Now we assume that there are four nodes which occur in pairs on 
opposite sides of the Fermi surface.
Partners in each pair produce equal and opposite Doppler shifts.
Summing over nodes we find
\begin{eqnarray}
G_0(0,\bm r)& =& \frac{-i}{N_F 2\pi^3v_F v_g^2}
\bigg[4p_0\Gamma_0+\pi (\alpha_1^2(\bm r)+\alpha_2^2(\bm r)
-2\Gamma_0^2)\bigg] \nonumber\\
  & \equiv &  \frac{\Gamma}{i\Gamma_0} 
\label{self}
\end{eqnarray} 
This result can easily be generalized to include more pairs of
nodes.
Equating the imaginary parts of (\ref{self}) yields
the self-consistent equation for the scattering rate
$\Gamma_0$, 
\begin{equation}
\Gamma_0=\frac{\pi^3}{2}\frac{N_F v_F v_g^2 \Gamma}
{p_0\Gamma_0-\frac{\pi}{2}\Gamma_0^2+\frac{\pi}{4}(\alpha_1^2(\bm r) + 
\alpha_2^2(\bm r))}.
\end{equation}
This equation describes how the scattering rate due to impurities
is modified in the presence of Doppler shifted quasiparticles.

As in 
Section II, we will assume that the magnetic field is parallel to the $xy$-plane
with an angle $\epsilon$ from the $x$ axis,
\be
{\bm H}=H(\cos\epsilon\ {\hat x}+\sin\epsilon\ {\hat y}).
\ee
The supercurrent is
\be
{\bm v}_s(\bm r)=\frac{1}{2mr}(-\sin\epsilon\cos\beta \hat{x}+\cos\epsilon\cos\beta\ \hat{y}+\sin\beta \hat{z}).
\ee
For now we will assume that all pairs of nodes are in the $xy$-plane at the
positions
\be
{\bm k_{F1}}&=&\pm k_F(\cos\phi_0\hat{x}-\sin\phi_0\hat{y})
\\
{\bm k_{F2}}&=&\pm k_F(\cos\phi_0\hat{x}+\sin\phi_0\hat{y}).
\ee
The angle $\phi_0$ is zero in the A phase of \pr\ (and the gap
function is doubly degenerate) and $\phi_0 \neq 0$ in the B phase.
This corresponds to the choice of the domain $(1,0,0)$ of the
A phase and the domain $(|\eta_1|,i|\eta_2|,0)$ of the B phase.  In each phase,
two other domains are possible and these will be discussed in the
next section.

The Doppler shifts  are
\begin{eqnarray}
\alpha_1(\bm r)&=&\pm {\bm v}_s(\bm r)\cdot{\bm k_F}_1\nonumber \\
& =& \pm \frac{k_F}{2mr}\cos\beta[-\sin\phi_0\cos\epsilon-\cos\phi_0\sin\epsilon] \\
\alpha_2(\bm r)&=&\pm {\bm v}_s(\bm r)\cdot{\bm k_F}_2\nonumber \\
& =& \pm \frac{k_F}{2mr}\cos\beta[\sin\phi_0\cos\epsilon-\cos\phi_0\sin\epsilon]
\end{eqnarray}

The following averages over the vortex unit cell will be useful:
\begin{eqnarray}
\left<\alpha_1^2(\bm r)+\alpha_2^2(\bm r)\right>_H&\sim & 
\frac{v_F^2}{R^2}\ln\left(\frac{R}{\xi_0}\right)
[\cos^2\phi_0\sin^2\epsilon
\nn\\
&&+\sin^2\phi_0\cos^2\epsilon]
\nn \\
\left<\alpha_1^2(\bm r)-\alpha_2^2(\bm r)\right>_H&\sim &\frac{v_F^2}{R^2}\ln\left(\frac{R}{\xi_0}\right)
\sin2\phi_0\sin2\epsilon.
\end{eqnarray}

\subsection{Density of states}
The DOS is given by Eq.\ \ref{dos}.  Using  (\ref{4.1.1}) and (\ref{self}) we find 
\be
N(0,{\bm r})&=&\frac{2\Gamma_0^2}{\pi^3 v_F v_g^2}
\bigg[\frac{p_0}{\Gamma_0}-\tan^{-1}\left(\frac{p_0}{\Gamma_0}\right)
\nn\\
&&+\frac{\alpha_1^2(\bm r)+\alpha_2^2(\bm r)}{2\Gamma_0^2}
\tan^{-1}\left(\frac{p_0}{\Gamma_0}\right)\bigg].
\ee
In zero magnetic field we retain our previous result for the impurity induced density of states\cite{Tayseer2007/2}, 
then the magnetic contribution is
\be
\delta N(0,{\bm r})&\approx&\frac{\alpha_1^2(\bm r)+\alpha_2^2(\bm r)}
{2\pi^2v_F v_g^2}
\ee
which depends upon the Doppler shifts to the same power 
as that of the frequency in the low frequency DOS in zero magnetic field,
$N(\omega) \sim \omega^2$ for superconductors with point nodes.
Taking the average over the vortex unit cell, we get
\begin{equation}
\label{density}
\frac{\left<\delta N(0,{\bm r})\right>_H}{N_F}\sim\frac{H}{H_{c2}}
\ln\left(\frac{H_{c2}}{H}\right)
[\cos^2\phi_0\sin^2\epsilon+\sin^2\phi_0\cos^2\epsilon].
\end{equation}
Thus we find that the DOS oscillates with
rotating magnetic field as $\cos 2 \epsilon$ and  is universal {\em i.e.} 
independent of the scattering rate.
\begin{figure}[ht]
\epsfysize=50mm
\epsfbox{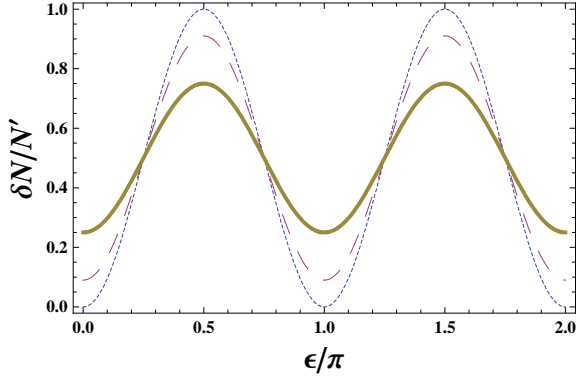}
\caption{(Color online) Oscillations in the density of states
(Eq.\ \ref{density}) for different values of
$\phi_0$ as a function of the field angle $\epsilon$. 
The dotted line is for $\phi_0=0$ (A phase), the dashed line
for $\phi_0=\arcsin(0.3)$ and the bold line for $\phi_0=\frac{\pi}{6}$. 
$N'\sim N_F\frac{H}{H_{c2}}
\ln\left(\frac{H_{c2}}{H}\right)$. }
\end{figure}

\subsection{Thermal Conductivity}
Beginning with Eq.\ \ref{kubo}, we divide the
volume of integration into parts centred around each node,
\be
\frac{\widetilde{\kappa}(0,{\bm r})}{T}&=&\frac{k_B^2}{3}\frac{\sum_{j=1}^4{\bm v}_F{\bm v}_F}{(2\pi)^3 v_F v_g^2}
\int_0^{2\pi}d\phi\int_0^{\pi}d\theta \sin\theta 
\nn\\
&&\int_0^{p_0}
dp\ p^2 \frac{\Gamma_0^2}{[(\alpha_j(\bm r)+p)^2+\Gamma_0^2]^2}
\ee
where the integration variable is again $p = 
 \sqrt{v_F^2 k_1^2 + v_g^2 (k_2^2 + k_3^2)}$.
The integrations yield
\begin{eqnarray}
\frac{\widetilde{\kappa}(0,{\bm r})}{T}&=&
\frac{k_B^2}{12\pi^2}\frac{\sum_{j=1}^4{\bm
v}_F{\bm v}_F}{v_F
v_g^2}\left[\frac{\Gamma_0^2+\alpha_j^2({\bm r})}{\Gamma_0}\right] \nn\\
&&\times\left[\frac{\pi}{2}-\tan^{-1}\frac{\alpha_j(\bm r)}
{\Gamma_0}-\frac{\alpha_j(\bm r)\Gamma_0}{\alpha_j^2(\bm r)+\Gamma_0^2}\right]
\end{eqnarray}
where again our previously derived expression for the residual conductivity in zero magnetic field\cite{Tayseer2007/2} is recovered.
The matrix ${\bm v}_F{\bm v}_F$ for one node is equal to the
contribution for the node on the opposite side of the Fermi surface,
but $\alpha_j(\bm r)$ changes sign at opposite nodes, therefore terms which are
odd in $\alpha_j(\bm r)$ will vanish.  The sum over nodes yields (keeping only
the magnetic part)
\begin{widetext}
\be
\frac{\delta\widetilde{\kappa}(0,\bm r)}{T}&=&\frac{k_B^2}{12 \pi}\frac{1}{v_F
v_g^2 \Gamma_0} \left[\alpha_1^2(\bm r)({\bm v}_F{\bm v}_F)_1
+\alpha_2^2(\bm r)({\bm
v}_F{\bm v}_F)_2\right]
\nn\\
&=& \frac{k_B^2}{12\pi}\frac{v_F}{ v_g^2\Gamma_0}\left(
  \begin{array}{ccc} (\alpha_1^2(\bm r)+\alpha_2^2(\bm r))\cos^2\phi_0&
  \frac{1}{2}(\alpha_2^2(\bm r)-\alpha_1^2(\bm r))\sin2\phi_0 & 0\\
  \frac{1}{2}(\alpha_2^2(\bm r)-\alpha_1^2(\bm r))\sin2\phi_0&
  (\alpha_1^2(\bm r)+\alpha_2^2(\bm r))\sin^2\phi_0 & 0 \\ 0& 0&0
\end{array} \right) . 
\ee
\end{widetext}
Finally we perform the average over the vortex unit cell,
\begin{widetext}
\be 
\left<\frac{\delta\widetilde{\kappa}(0,\bm r)}{T}\right>_H\sim k_B^2\frac{v_F\Delta_0^2}{v_g^2\Gamma_0}\frac{H}{H_{c2}} \ln\left(\frac{H_{c2}}{H}\right) \left(
\begin{array}{ccc}
\cos^2\phi_0[\cos^2\phi_0\sin^2\epsilon+\sin^2\phi_0\cos^2\epsilon]&
-\frac{1}{4} \sin2\phi_0\sin2\epsilon&0\\ -\frac{1}{4}
\sin2\phi_0\sin2\epsilon&\sin^2\phi_0[\cos^2\phi_0\sin^2\epsilon+\sin^2\phi_0\cos^2\epsilon]
&0\\ 0& 0&0
\end{array} \right) 
\nn\\
\label{n1n20}
\ee
\end{widetext}
where $\Delta_0^2=\frac{v_F^2}{\xi_0^2}\sim N_F v_F v_g^2$.
The A phase of \pr\ corresponds to $\phi_0 =0$, and the only component
of the thermal conductivity which is non-vanishing is $\kappa_{xx}
\sim \sin^2\epsilon$.

\subsubsection*{Other domains}
The phase $D_2(E)$ has 
two other nodal configurations,\cite{note3}
which may be found by applying the operation $C_3$ on the components
$(|\eta_1|,i|\eta_2|,0)$ or directly on the gap function.
The second domain we consider is when the nodes are in the
$k_xk_z$-plane. Then the B phase has order parameter components
$(i|\eta_2|,0,|\eta_1|)$ and the A phase has components $(0,0,1)$.
Then the positions of the nodes are
\be
\bm k_{F1}&=&\pm k_F(-\sin{\phi_0}\hat{x}+\cos{\phi_0}\hat{z})
\\
\bm k_{F2}&=&\pm k_F(\sin{\phi_0}\hat{x}+\cos{\phi_0}\hat{z})
\ee
and the Doppler shifts are
\begin{eqnarray}
\alpha_1(\bm r)&=& \pm \frac{k_F}{2mr}[\sin\phi_0\sin\epsilon\cos\beta+\cos\phi_0\sin\beta]
\\
\alpha_2(\bm r)&=&\pm\frac{k_F}{2mr}[-\sin\phi_0\sin\epsilon\cos\beta+\cos\phi_0\sin\beta]
\end{eqnarray}
In the average over the vortex unit cell $\left<\alpha_1^2(\bm r)-\alpha_2^2(\bm r)\right>_H$ vanishes, 
and 
\begin{equation}
\left<\alpha_1^2(\bm r)+\alpha_2^2(\bm r)\right>_H\sim\frac{v_F^2}{R^2}\ln\left(\frac{R}{\xi_0}\right)
[\sin^2\phi_0\sin^2\epsilon+\cos^2\phi_0].
\end{equation}
Then the B phase thermal conductivity is
\begin{eqnarray}
\left<\frac{\delta\widetilde{\kappa}(0,\bm r)}{T}\right>_H 
&\sim&  k_B^2\frac{v_F\Delta_0^2}{v_g^2\Gamma_0}\frac{H}{H_{c2}}
\ln\left(\frac{H_{c2}}{H}\right)
[\sin^2\phi_0\sin^2\epsilon
\nn\\
&&+\cos^2\phi_0]
\left( \begin{array}{ccc}
\sin^2\phi_0& 0&
0 \\
 0 & 0 &0\\
0& 0&\cos^2\phi_0 
\end{array} \right)
\label{n10n2}
\end{eqnarray}
and the A phase thermal conductivity is $\kappa_{zz} \sim$ constant.
\begin{figure}[ht]
\epsfysize=50mm
\epsfbox{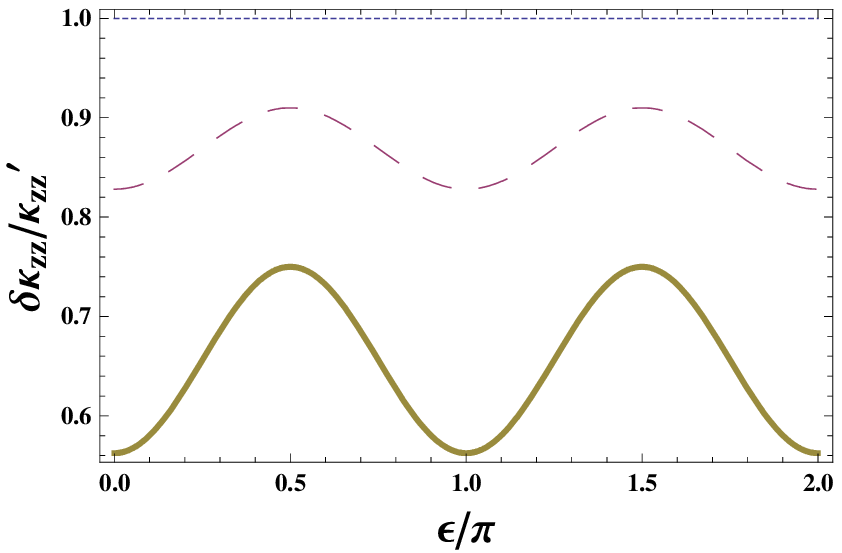}
\caption{(Color online) $\kappa_{zz}$
 (Eq.\ \ref{n10n2}) for different values of
$\phi_0$ as a function of the field angle $\epsilon$. 
The dotted line is for $\phi_0=0$ (A phase), the dashed line
for $\phi_0=\arcsin(0.3)$ and the bold one is for $\phi_0=\frac{\pi}{6}$.
$(\kappa_{zz}'/T)\sim k_B^2\frac{v_F\Delta_0^2}{v_g^2\Gamma_0}\frac{H}{H_{c2}}
\ln\left(\frac{H_{c2}}{H}\right) $. }
\end{figure} 
\begin{figure}[ht]
\epsfysize=50mm
\epsfbox{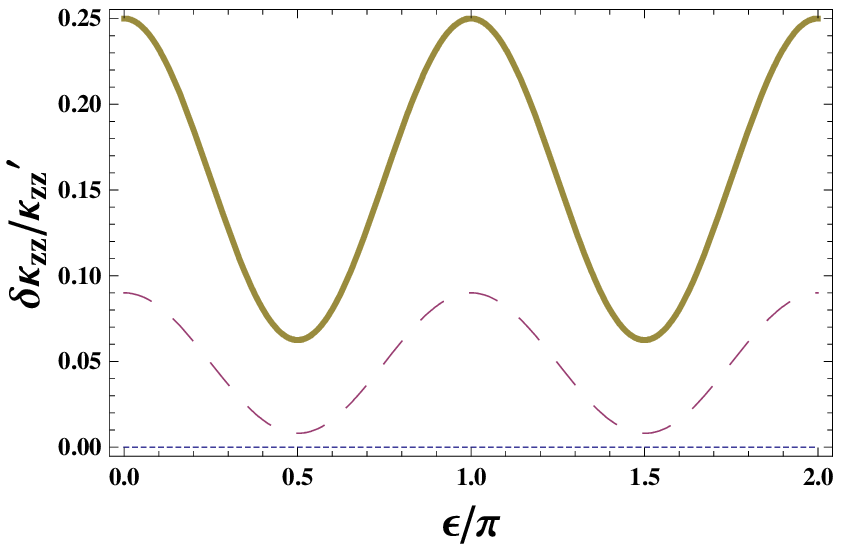}
\caption{(Color online) $\kappa_{zz}$
(Eq.\ \ref{0n1n2}) for different values of
$\phi_0$ as a function of the field angle $\epsilon$. 
The dotted line is for $\phi_0=0$ (A phase), the dashed line
for $\phi_0=\arcsin(0.3)$ and the bold one is for $\phi_0=\frac{\pi}{6}$.
$(\kappa_{zz}'/T)\sim k_B^2\frac{v_F\Delta_0^2}{v_g^2\Gamma_0}\frac{H}{H_{c2}}
\ln\left(\frac{H_{c2}}{H}\right) $.}
\end{figure} 
In the third  domain the nodes are found in the  $k_yk_z$-plane
\be
{\bm k_{F1}}&=&\pm k_F(\cos\phi_0\hat{y}-\sin{\phi_0}\hat{z})
\\
{\bm k_{F2}}&=&\pm k_F(\cos\phi_0\hat{y}+\sin{\phi_0}\hat{z})
\ee
In the average over the vortex unit cell, 
$\alpha_1^2(\bm r)-\alpha_2^2(\bm r)$ again vanishes, 
and we find
\begin{equation}
\left<\alpha_1^2(\bm r)+\alpha_2^2(\bm r)\right>_H\sim\frac{v_F^2}{R^2}
\ln\left(\frac{H_{c2}}{H}\right)
[\sin^2\phi_0+\cos^2\phi_0\cos^2\epsilon] .
\end{equation}
Then the B phase thermal conductivity is
\be
\left<\frac{\delta\widetilde{\kappa}(0,\bm r)}{T}\right>_H
&\sim& k_B^2\frac{v_F\Delta_0^2}{v_g^2\Gamma_0}\frac{H}{H_{c2}}
\ln\left(\frac{H_{c2}}{H}\right)
\nn\\
&&\times[\sin^2\phi_0+\cos^2\phi_0\cos^2\epsilon]
\nn\\
&&\times\left( \begin{array}{ccc}
0& 0&0 \\
 0 & \cos^2\phi_0 &0\\
0& 0&\sin^2\phi_0 
\end{array} \right)
\label{0n1n2}
\ee
and the only non-vanishing component in the A phase is $\kappa_{yy} \sim \cos 2\epsilon$.

\subsubsection*{Domain averaging}
In real situations, one may expect that either a single domain will form
either because of sample shape or applied strains or fields, or
that all three domains will be present.  If all
three domains are present then detailed knowledge of the
domain structure is required to calculate the conductivity.  Lacking 
such knowledge, we consider two limiting cases: {\em i)} serial domains and
{\em ii)} parallel domains.  
When the domains are in series  the conductivity is
$\widetilde{\kappa} = \left(\widetilde{\kappa}_1^{-1} + \widetilde{\kappa}_2 ^{-1} + \widetilde{\kappa}_3^{-1}\right)^{-1}$
which vanishes in all components.
When the domains are in parallel 
the three conductivities are simply added:
\begin{widetext}
\begin{equation}
 \left<\frac{\delta\widetilde{\kappa}(0,\bm r)}{T}\right>_H\sim 
\left(\begin{array}{ccc}
\sin^2\epsilon (1-\frac{3}{4} \sin^2 2\phi_0) + \frac{1}{2}\sin^2 2\phi_0 &
-\frac{1}{4}\sin^2\phi_0\sin 2\epsilon & 0 \\
-\frac{1}{4}\sin^2 2\phi_0 \sin 2\epsilon & 
\cos^2\epsilon (1-\frac{3}{4} \sin^2 2\phi_0) + \frac{1}{2}\sin^2 2\phi_0&0 \\
0 &0 &  1 - \sin^2\phi_0\cos^2\phi_0 \end{array} \right)
\label{n1n20average}
\end{equation}
\end{widetext}
for the B phase, while the result for the A phase ($\phi_0=0$) is
\begin{equation}
 \left<\frac{\delta\widetilde{\kappa}(0,\bm r)}{T}\right>_H\sim 
\left(\begin{array}{ccc}
\sin^2 \epsilon & 0 & 0 \\
0 & \cos^2 \epsilon & 0 \\
0 & 0 & 1 \end{array}\right)
\label{100average}
\end{equation}
Also, the domain averaged density of states is constant (has no oscillations).
\subsubsection*{Other nodal configurations}
According to Table I of Ref.\ \onlinecite{Sergienko2004}\cite{note2} there are other
nodal configurations corresponding to
other superconducting phases which should be considered.
Superconducting phases with cusp point nodes  in tetrahedral superconductors
are summarised in Table~\ref{tableI}.
\begin{table}[ht]
\begin{tabular}{l|l|l|l}
\hline
Nodes & Symmetry & IR & channel\\
\hline
4 nodes $[0,\cos\phi_0,\pm\sin\phi_0]$ & $D_2(E)$& $T_u$ & triplet\\
\hline
2 nodes $[0,0,1]$ & $D_2(C_2)\times{\cal K}$, $C_3(E)$ & $T_u$ & triplet\\
6 nodes $[0,0,1]$ & $C_3\times{\cal K}$, $C_2'(E)$, ${\cal K}$, $E$ & $T_g$& singlet\\
\hline
2 nodes $[1,1,1]$ & $C_2(E)\times{\cal K}$ & $T_u$ & triplet\\
8 nodes $[1,1,1]$ & $T(D_2)$, $D_2\times{\cal K}$, $D_2$ &$E_g$ & singlet \\
8 nodes $[1,1,1]$             &  $T(D_2)$ & $E_u$ & triplet \\
\hline
6 nodes $[0,0,1]$ and  & & \\
2 nodes $[1,1,1]$ & $C_3(E)$ & $T_g$ & singlet\\
\hline
\end{tabular}
\caption{Cusp point nodal configurations, their associated symmetries,
order parameters labeled by irreducible representation (IR) and
pairing channels for a tetrahedral superconductor 
(after Ref.\ \onlinecite{Sergienko2004}).  
\label{tableI}}
\end{table}

For {\bf eight point nodes in the $[111]$ directions}, the
thermal conductivity is
\begin{equation}
\left<\frac{\delta\widetilde{\kappa}(0,\bm r)}{T}\right>_H \sim 
\left(\begin{array}{ccc} 
2 & - \sin 2 \epsilon & 0 \\
-\sin 2\epsilon & 2 & 0 \\
0 & 0 & 2 \end{array} \right)
\label{111average}
\end{equation}

The thermal conductivity for {\bf two point nodes in the
$[111]$ directions} is
\begin{equation}
\left<\frac{\delta\widetilde{\kappa}(0,\bm r)}{T}\right>_H\sim
\left(1-\frac{\sin 2 \epsilon}{2}\right)\left(
\begin{array}{ccc}
1 & 1 & 1 \\
1 & 1 & 1 \\
1 & 1 & 1\end{array}\right) .
\label{111}
\end{equation}
Such a phase has three other domains, each with a single pair of nodes
in the directions $[1-1-1]$, $[-11-1]$ and $[-1-11]$; 
the parallel domain averaged conductivity is 
equivalent to (\ref{111average}).

The thermal conductivity 
for  {\bf six point nodes in the $[100]$ directions}
is given by (\ref{100average}).

The thermal conductivity for {\bf six point nodes
in the $[100]$ directions and two point nodes in the 
$[111]$ directions} is given by the sum 
of (\ref{100average}) and (\ref{111}).  Such a phase has four domains;
the parallel domain averaged conductivity is given
by the sum of (\ref{100average}) and (\ref{111average}).

Eqs.\ \ref{n1n20}, \ref{n10n2} and \ref{0n1n2}-\ref{111}
are summarized in Table \ref{kappa}.
In all cases, the highest harmonics which appear are two-fold oscillations,
stemming from the fact that contributions from pairs of nodes 
are additive and 
proportional to the square of the Doppler shift. 
\begin{table}[ht]
\begin{tabular}{l|ccccc}
\hline
Nodes & $\kappa_{xx}$ & $\kappa_{yy}$ & $\kappa_{xy}$ & $\kappa_{xz,yz}$ &
$\kappa_{zz}$ \\
\hline
4 nodes  $[\cos\phi_0,\pm\sin\phi_0,0]$  & c & c & s & 0 & 0\\
4 nodes  $[\pm\sin\phi_0,0,\cos\phi_0]$  & c & 0 & 0 & 0 & c\\
4 nodes  $[0,\cos\phi_0,\pm \sin\phi_0]$ & 0 & c & 0 & 0 & c\\
domain average  & c&c&s&0&1\\
\hline
2 nodes $[1,0,0]$ & c & 0 & 0 & 0 & 0\\
2 nodes $[0,1,0]$ & 0 & c & 0 & 0 & 0\\
2 nodes $[0,0,1]$ & 0 & 0 & 0 & 0 & 1\\
domain average/ 6 nodes $[1,0,0]$ & c & c & 0 & 0 & 1\\
\hline
2 nodes $[1,1,1]$ & s & s & s & s & s \\
domain average/ 8 nodes $[1,1,1]$ & 1 & 1 & s & 0 & 1\\
\hline
6 nodes $[1,0,0]$ and 2 nodes $[1,1,1]$ & c+s & c+s & s & s & s\\
domain average & c & c & s & 0 & 1 \\
\hline
\end{tabular}
\caption{Oscillatory contributions to the thermal conductivity 
with a field rotating in the $xy$ plane for various
nodal configurations. `s' stands for $\sin 2\epsilon$,
`c' stands for $\cos 2\epsilon$, `1' stands for no
oscillations and `0' means that the component vanishes.
$\epsilon$ is the angle of the field with respect to the $x$ axis.
\label{kappa}}
\end{table}

\section{Discussion}
So far there has only been one report of thermal conductivity in 
a rotating magnetic field, namely the results by Izawa {\em et al.},\cite{Izawa2003}
who measured $\kappa_{zz}$ and found four-fold oscillations
near $H_{c2}$ and a sharp transition to two-fold oscillations at a lower field.
We do not obtain four-fold oscillations for any of the 
point node configurations we considered and so we conclude
that the formalism we have used is inapplicable in
large magnetic fields.  One possible source of 
error is that we have omitted contributions from
quasi-particle states localised in vortex cores, and that
these states may dominate the oscillatory contribution to
the density of states as the field increases and the vortices
become closer together.  Another possibility is higher order in
$\alpha(\bm r)$ (the Doppler shift)
contributions become important as the field is increased.
We do not obtain four-fold oscillations simply because we did
not retain contributions to the density of states
and thermal conductivity for powers of $\alpha(\bm r)$ higher than two.
In any case, unlike $d_{x^2-y^2}$ line node superconductors,
the four-fold oscillations reported in 
Ref.\ \onlinecite{Izawa2003} are not related in any simple
way to the nodal structure of \pr.

Our results may be applicable to the lower field thermal 
conductivity 
measurements
in which two-fold oscillations are found.  The lower
inset of Fig.\ 1 of Ref.\ \onlinecite{Izawa2003} shows a nearly
linear dependence of $\kappa_{zz}$ on $H$, in rough agreement 
with $H \log H$ dependence expected for point nodes
(see Eq.\ \ref{n1n20}).
Fig.\ 2b) of 
Ref.\ \onlinecite{Izawa2003} clearly shows two-fold oscillations
of the form $\kappa_{zz} \sim \cos 2\epsilon$ and {\em not} $\sin 2\epsilon$.
This indicates that the most likely superconducting
phase of \pr\ is $D_2(E)$
which belongs to the three dimensional
order parameter $T_u$ and that a single
domain with order parameter components $(0,|\eta_1|,i|\eta_2|)$ or
$(i|\eta_2|,0,|\eta_1|)$ was measured in Ref.\ \onlinecite{Izawa2003}.
We note that this phase agrees with various properties observed
in other experiments, including triplet pairing,\cite{Higemoto2007}
broken time reversal symmetry\cite{Aoki2003} and broken $C_3$ symmetry.\cite{Huxley2004,Mukherjee2006}

Sakakibara et al.\cite{Sakakibara2007} measured the field-angle
dependent specific heat 
and found four-fold oscillations, but unlike Izawa et al., 
no four-fold to two-fold transition
was observed.
In this case we must also attribute the four-fold oscillations 
to corrections beyond the semi-classical methods we have used.
Sakakibara et al.\cite{Sakakibara2007} verified that, in their 
set-up, superconductivity has no preferred orientation, implying that their
results are domain averaged.  Since two-fold oscillations are not
observed at all, these results may be consistent with any of the domain 
averaged
configurations shown in Table \ref{kappa}. 
 
Similarly to computational issues pertaining to 
domain averaging, the thermal conductivity must also be
averaged over the vortex lattice.
In all of our calculations we
performed the vortex average as a simple areal average over a plane
perpendicular to a vortex, as shown in Eq.\ \ref{average}. 
This procedure is appropriate when the
heat current is parallel to the vortices.  For currents in
other directions a different averaging procedure should be used,
which results in a more complicated
field dependence of the oscillation amplitudes than
what we have shown here.
The correct procedure is an average of 
$\kappa$ over paths through the vortex lattice,
which is in fact more involved than the series average
$\langle \kappa^{-1} \rangle^{-1}$ described in Ref.\ \onlinecite{Kubert1998/2}.
This means that
the vortex averaging calculation for any
in-plane component of the conductivity will vary with the
field angle, producing oscillations $\sim \cos 2\epsilon$
which are unrelated to nodes and 
which will dominate over any nodal contributions.\cite{Matsuda2006,Maki1967}
Thus observations of oscillations $\sim \cos 2 \epsilon$ in
$\kappa_{xx}$, $\kappa_{yy}$ or in off-diagonal components of $\kappa$ 
measured with an in-plane current should not be interpreted
as evidence of nodes.

For $\kappa_{zz}$, $\kappa_{xz}$ and $\kappa_{yz}$, when the 
current is perpendicular to vortices, there
will not be oscillations due to the vortex averaging. 
If observed, oscillations may therefore be attributed to nodes.
A small oscillatory contribution will arise from
mixing with the other components of $\kappa$
via the vortex averaging procedure (which 
in general does involve averaging $\kappa^{-1}$)
but we expect it to be small compared to the
oscillations originating from nodes.

\section{Summary}
We have reviewed previous works concerning 
field-angle dependent DOS and thermal conductivity for line 
node superconductors using a semi-classical method, and
applied the same method to point node superconductors.
This method  neglects vortex localised quasi-particles and
retains only the contribution from extended, nodal quasiparticles
to the density of states.
Clearly there are limitations to this approach; in particular
it cannot be expected to produce an accurate estimate
of the total low-energy density of states in point node superconductors.
However it may be a reasonable way to estimate the
field-angle dependent oscillatory component of the
density of states and related quantities
for fields $H_{c1} \leq H \ll H_{c2}$.
We find that in point node superconductors there is no difference
between the clean and dirty limits, unlike in line node superconductors
in which the different limits produce significantly different 
expressions for the oscillatory part of the thermal conductivity.
Considering all possible configurations of point nodes in a 
tetrahedral superconductor, we find that 
the superconducting phase $D_2(E)$, which we
previously proposed based on other experimental evidence,
best accounts for field-angle dependent
oscillations in the thermal conductivity 
of \pr.

\acknowledgements

We thank Ilya Vekhter for helpful discussions.  We gratefully acknowledge
the hospitality of the University of Waterloo, where this
work was completed.  This work was supported by
NSERC of Canada.

\end{document}